\documentstyle[preprint,revtex,eqsecnum]{aps}

\begin{document}
\preprint{hep-ph 9302287}
\draft
\begin{title}
Possibilities beyond 3-3-1 models
\end{title}
\author{ Vicente Pleitez}
\begin{instit}
Instituto de F\'\i sica Te\'orica\\
Universidade Estadual Paulista\\
Rua Pamplona, 145\\
01405-900-- S\~ao Paulo, SP\\
Brazil
\end{instit}
\begin{abstract}
We consider generalizations of the standard model (SM) which are
based on the gauge symmetry $SU(n)_c\otimes SU(m)_L\otimes U(1)_N$.
Although the most interesting possibilities occur when $n=3$, we will
consider also
the cases $n=4,5$ both with $m=3,4$. Models with
left-right symmetry, horizontal symmetries and the
possible embedding in a larger group (grand unification scenarios) are
briefly discussed.
\end{abstract}
\pacs{PACS numbers:           12.15.Cc; 12.10.Dm}

\section{Introduction}
\label{sec:int}
Recently, a class of $SU(3)_c\otimes SU(3)_L\otimes
U(1)_N$ models was considered in Refs.~\cite{pp,mpp,pt} with
several representation contents (called 3-3-1 models by Frampton
Ref.~\cite{pp}.) Such models are
anomaly free only if there is equal number of triplets and antitriplets
(considering the color degrees of freedom), and furthermore requiring
the sum of all fermion charges to vanish. Two of the three quark
generations transform identically and one generation, it does not
matter which~\cite{pp2}, transforms in a different representation of
$SU(3)_L \otimes U(1)_N$. This means that in these models, which are
undistinguishable from the standard model at low energies, in order
to cancel anomalies, the number of families ($N_f$) must be divisible
by the number of color degrees of freedom ($n$). Hence the simplest
alternative is $n=N_f$. Another interesting feature of these models is
that the weak mixing
angle of the standard model has an upper limit. For instance, in the
model of Refs.~\cite{pp} $\sin^2\theta_W$ has to be smaller than
1/4. Therefore, it is possible to compute an upper limit to the mass
scale of the $SU(3)$ breaking of about $1.7$ TeV~\cite{ng}. This makes
3-3-1 models interesting possibilities for the physics beyond the
standard model. Hence, it is also interesting to study how the
basic ideas of this sort of models can be generalized.

Here we want to generalize this models in several ways: by
expanding the color degrees of freedom $(n)$ and the electroweak
sector $(m)$, i.e., we will consider models based on the gauge
symmetry
\begin{equation}
SU(n)_c\otimes SU(m)_L\otimes U(1)_N.
\label{g}
\end{equation}
In all these extensions the anomaly cancellation occurs among all
generations together, and not
generation per generation. Models with left-right symmetry and/or
with horizontal symmetry are also considered. We discuss too
possible grand unified theories in which some of these models may be
embedded.

We will use the criterion that the values for $m$ are determined by
the leptonic sector. In the color sector for
simplicity, in addition to the usual case of $n=3$, we will
comment the cases $n=4,5$. These extensions have been considered in
the context of the $SU(2)_L\otimes U(1)_Y$ model~\cite{f,fh}.

This work is organized as follows. In Sec.\ref{sec:nm1} we will
consider models with $n=3$, $m=3,4$ (Sec.\ref{subsec:341}). There we
will also discuss the cases when $n=4,5$ (Sec.\ref{subsec:n45}). In
Sec.\ref{sec:lrhs} we will give general features of the extensions with
left-right symmetry (Sec.\ref{subsec:lr}) and with horizontal symmetries
(Sec.\ref{subsec:ho}). We also consider (Sec.\ref{subsec:su6}) possible
embedding in $SU(6)$. The last section is devoted to our conclusions.

\section{Models with extended color and electroweak sectors}
\label{sec:nm1}
\subsection{Models with extended electroweak sector}
\label{subsec:341}
First, let us consider the $n=3$ models. In this case, $m=2$ gives
the standard model. The case of $m=3$ is possible
with three leptons belonging to the fundamental representation of
$SU(3)_L$. As this kind of models has been already considered in
detail in Refs.~\cite{pp,mpp,pt} here we will treat them
briefly. Consider
leptons transforming as triplets $({\bf3},0)$ under
$SU(3)_L\otimes U(1)_N$~\cite{pp}:
$(\nu_a,l_a,l^c_a)^T_L$, where $a=1,2,3$ is the family index and
$l^c_a$ are the charge conjugate
fields. In this model there are, besides the usual
quarks of the standard model, three exotic ones with charges
$\frac{5}{3}$ (one) and $-\frac{4}{3}$ (two). It means that there are
nine quarks, each one with the three usual color degrees of freedom.
It is also possible the following representation content
$(\nu^c_a,\nu_a,l_a)^T_L\sim ({\bf3}^*,-1/3)$,
and it is necessary to introduce new quarks with the same
charge of
the known quarks. The model has also 9 quarks, four of them with charge
$2/3$ and five  with charge $-1/3$~\cite{mpp,s}. Let us write down
explicitly the quark content since this model will be considered in
Sec.~\ref{subsec:su6}. The first and second ``generations'' are in
triplets $({\bf3},0)$,  and the third one in an antitriplet
$({\bf3}^*,1/3)$~\cite{fn1}:
\begin{equation}
Q_{iL}=\left(\begin{array}{c}
u_i \\ d_i \\ d'_i\end{array}\right)_L\sim ({\bf3},0),\, i=1,2;\,
Q_{3L}=\left(\begin{array}{c}
t'\\ t \\ b\end{array}\right)_L\sim ({\bf3}^*,1/3).
\label{modelo1}
\end{equation}
All right-handed charged fermions are taken to be $SU(3)$ singlets.
The representations above are in terms of symmetry eigenstates.

Next, we consider an example of a 3-4-1 model in which the electric
charge operator is defined as
\begin{equation}
Q=\frac{1}{2}(\lambda_3-\frac{1}{\sqrt3}\lambda_8-\frac{2}{3}{\sqrt6}
\lambda_{15})+N,
\label{q}
\end{equation}
where the $\lambda$-matrices are a slightly modified version of the
usual ones~\cite{da},
\[\lambda_3=diag(1,-1,0,0),\;\lambda_8=(\frac{1}{\sqrt{3}})diag(1,1,-2,0),
\;\lambda_{15}=(\frac{1}{\sqrt{6}})diag(1,1,1,-3).\]

Leptons transform as $({\bf1},{\bf4},0)$,
two of the three quark families, say $Q_{iL},\,i=2,3$, transform as
$({\bf3},{\bf4}^*,-1/3)$,
and one family, $Q_{1L}$, transforms as $({\bf3},{\bf4},+2/3)$
\begin{equation}
\psi_{aL}=\left(\begin{array}{c}
\nu_a\\
l_a\\
\nu^c_a\\
l^c_a
\end{array}\right)_L, \qquad Q_{1L} = \left(
\begin{array}{c}
u_1\\
d_1\\
u'_1\\
J
\end{array}\right)_L, \qquad Q_{iL} = \left(
\begin{array}{c}
j_i\\
d'_i\\
u_i\\
d_i
\end{array}\right)_L,
\label{lq}
\end{equation}
where $u'_1$ and $J$ are new quarks with charge $+2/3$
and $+5/3$ respectively;
$j_i$ and $d'_i$, $i=2,3$ are new quarks with charge $-4/3$ and
$-1/3$ respectively. We remind that in Eq.~(\ref{lq})
all fields are still symmetry eigenstates.
Right-handed quarks transform as singlets under $SU(4)$.

A model with $SU(4)_L$ symmetry and leptons transforming as in
Eq.~(\ref{lq}) was proposed by Voloshin some years
ago~\cite{voloshin}. In this context it can be
possible to understand the existence of neutrinos with large magnetic
moment and small mass. However in Ref.~\cite{voloshin} it was not
considered the quark sector.

Quark masses are generated by introducing the following Higgs
$SU(4)_L\otimes U(1)_N$ multiplets:
$\chi\sim({\bf4},+1),\rho\sim({\bf4},+1/3)$, $ \eta$ and
$\eta'\sim({\bf4},0)$.

In order to obtain massive charged leptons it is necessary to
introduce a $({\bf10}^*,0)$ Higgs multiplet, because the lepton mass
term transforms as
$\bar \psi^c_L\psi_L\sim ({\bf6}_A\oplus {\bf10}_S)$.
The ${\bf6}_A$ will leave some leptons massless and some others mass
degenerate. Therefore we will choose $H={\bf10}_S$.
Neutrinos remain massless at least at tree level but the charged
leptons gain mass. The corresponding VEVs are the following
$\langle\eta\rangle=(v,0,0,0)$, $\langle\rho\rangle=(0,w,0,0)$,
$\langle\eta'\rangle=(0,0,v',0)$,
$\langle\chi\rangle=(0,0,0,u)$,
and $\langle H\rangle_{42}=v''$ for the decuplet. In this way we have
that the symmetry breaking of the $SU(4)_L\otimes U(1)_N$ group down
to $SU(3)_L\otimes
U(1)_{N'}$ is induced by the $\chi$ Higgs. The $SU(3)_L\otimes U(1)_{N'}$
symmetry is broken down into $U(1)_{em}$ by the $\rho,\eta$, $\eta'$
and $H$ Higgs. As in the model I of Ref.~\cite{mpp}, it is necessary
to introduce some discrete symmetries which
ensure that the Higgs fields give a
quark mass matrix in the charge $-1/3$ and  $2/3$ sectors of the
tensor product form in order to avoid general mixing among
quarks of the same charge. In this case the quark mass matrices can
be diagonalized with unitary matrices which are themselves tensor
product of unitary matrices.

In fact, we have the symmetry breaking pattern, including the $SU(3)$
of color,
\begin{equation}\begin{array}{c}
SU(3)_c \otimes SU(4)_L \otimes U(1)_N \\
\downarrow \langle \chi \rangle\\
SU(3)_c \otimes SU(3)_L \otimes U(1)_{N'}\\
\downarrow \langle \eta'\rangle\\
SU(3)_c\otimes SU(2)_L \otimes U(1)_{N''}\\
\downarrow \langle x \rangle\\
SU(3)_c \otimes U(1)_{em}
\end{array} \label{sbp} \end{equation}
where $\langle x\rangle$ means $\langle \rho\rangle$, $\langle
\eta\rangle$, $\langle H\rangle$~\cite{ppx}.

The electroweak gauge bosons of this theory consist of a
${\bf15}$ $W^i_\mu$, $i=1,...,15$ associated with $SU(4)_L$ and
a singlet $B_\mu$ associated with $U(1)_N$.

There are four neutral bosons: a massless $\gamma$ and three massive
ones: $Z,Z',Z''$. The lightest one, say the $Z$, corresponds to the
Weinberg-Salam neutral boson. Assuming the approximation
$u\gg v'\gg v,v'',w$ the extra neutral bosons,
say $Z',Z''$, have masses which depend mainly on $u,v'$.

Concerning the charged vector bosons, as in the model of
Refs.~\cite{pp} there are doubly charged vector bosons and there are
doublets of $SU(2)$
$(X^+_\mu,X^0)$ and $(\bar X^0_\mu,X^-_\mu)$
which produce interactions like $\bar
\nu^c_{aL}\gamma^\mu l_{aL}X^+_\mu$ and
$\bar\nu^c_{aL}\gamma^\mu\nu_{aL} X^0_\mu$, as in
model I of Ref.~\cite{mpp}. We have also the
$V^\pm$ vector bosons with interactions like $\bar\nu_{aL}\gamma^\mu
l^c_{aL}V^-$.

\subsection{$n=4,5$ models}
\label{subsec:n45}
Next, we consider $n=4,5$ models. Although the $SU(3)_c$ gauge
symmetry is the best candidate for the theory of the strong
interactions, there is no fundamental reason
why the colored gauge group must be $SU(3)_c$. In fact, it is
possible to consider other Lie groups. In general we have the
possibilities $SU(n),\,n\geq 3$~\cite{so}.

In particular, models in which quarks transform
under the fundamental representations of $SU(4)_c$ and $SU(5)_c$ were
considered in Refs.~\cite{f} and \cite{fh} respectively. These models
preserve the experimental consistency of the standard model at low
energies. For instance, in the $SU(5)_c\otimes SU(m)_L\otimes U(1)_N$
model a Higgs
field transforming as the ${\bf10}$ representation of $SU(5)_c$
breaks the symmetry as follows~\cite{fh}
\begin{equation}
\begin{array}{c}
SU(5)_c \otimes SU(m)_L \otimes U(1)_N\\
\downarrow \langle {\bf10} \rangle\\
SU(3)_c \otimes SU(2)'\otimes SU(m)_L\otimes U(1)_N.
\end{array} \label{sbp3} \end{equation}
Later the electroweak symmetry will be broken and the remaining
symmetry will be $SU(3)_c\otimes SU(2)'\otimes U(1)_{em}$ as in the
models with $m=3,4$ considered above. Notice that, due to the
relation between the color degrees of freedom and the number of
families, it is necessary to introduce four and five families
for $n=4$ and $n=5$ respectively.

\section{Other possible extensions}
\label{sec:lrhs}
Other possibilities are, models with left-right symmetry in the
electroweak sector
$SU(n)_c\otimes SU(m)_L\otimes SU(m)_R\otimes U(1)_N$
and also models with horizontal symmetries $G_H$ i.e.,
$SU(n)_c\otimes SU(m)_L\otimes U(1)_N\otimes G_H$.

\subsection{Left-right symmetries}
\label{subsec:lr}
In models with left-right symmetry the $V-A$ structure of weak
interactions is related to the mass difference between the left- and
right- gauge bosons, $W^\pm_L$ and $W^\pm_R$, respectively, as a
result of the spontaneous symmetry breaking~\cite{gs}.

This sort of models is easily implemented in the 3-3-1 context by
adding a new lepton $E$. For instance, in models with
left-handed leptons transforming as $(\nu_a,l^-_a,E^+_a)^T_L$ the
right-handed triplet is $(\nu_a,l^-_a,E^+_a)^T_R$. In the quark sector, the
left-handed components are as in Ref.~\cite{pp} and similarly the
right-handed components in such a way
that anomalies cancel in each chiral sector. Explicitly, the charge
operator is defined as
\begin{equation}
Q=I_{3L}+I_{3R}+N
\label{colr}
\end{equation}
where $I_{3L(3R)}$ are of the form $(1/2)(\lambda_3-\sqrt{3}\lambda_8)$.
The Higgs multiplet $({\bf3},{\bf3}^*,0)$ and its conjugate give
mass to all fermions but in order to complete the symmetry breaking
it is necessary to add more Higgs multiplets. Details will be given
elsewhere.

\subsection{Horizontal Symmetries}
\label{subsec:ho}
For example,
if $G_H=SU(2)_H$ there are no additional conditions to
cancel anomalies since $SU(2)$ is a safe group. For instance, with
$n=3$ the three generations can transform in the adjoint
representation of $SU(2)_H$. Another possibility is $G_H=SU(3)_H$
with the three generations in the fundamental representation. In this
case it is necessary to introduce right-handed neutrinos. To avoid
anomalies, for example in the models of Refs.~\cite{pp}, we can choose
either i) all left-handed fermions are in ${\bf3}$
and the right-handed ones in ${\bf3}^*$ , or ii) the left-handed components
of the charge 2/3 and the right-handed components of the charge $-1/3$
transform as a ${\bf3}$, while the right-handed components of the
charge 2/3 and the left-handed components of the charge $-1/3$ quarks
transform as ${\bf3}^*$. Similarly in the
leptonic sector~\cite{gelmini}. The left-handed (right-handed)
components of the exotic quarks transform as ${\bf3}$ (${\bf3}^*$).
The consequences of these horizontal symmetries in 3-3-1 models on
the mass spectra of quarks and leptons deserve detailed studies.

\subsection{Embedding in $SU(6)$}
\label{subsec:su6}
There are also the grand unified extensions of all the possibilities
we have treated above.

For example, it is possible to embed
the 3-3-1 models in a simple group as $SU(6)$~\cite{gp}.
The group $SU(3)_c\otimes
SU(3)_L\otimes U(1)_N$ has rank 5 and it is a subgroup of
$SU(6)$. In the last group, it has been shown that the anomalies,
${\cal A}$, of ${\bf15}$ and ${\bf6}^*$ are such that ${\cal
A}({\bf15})=-2{\cal A}({\bf6}^*)$~\cite{bg}. Then, pairs of
${\bf15}$ and ${\bf15}^*$; ${\bf6}^*$ and ${\bf6}$~\cite{gg} and,
finally one ${\bf15}$ and two ${\bf6}^*$ are the smallest anomaly
free irreducible representations in $SU(6)$. On the other hand, the
representation ${\bf20}$ is safe.

Just as an example, let us consider the $SU(6)$ symmetry which is
a possible unified theory for the $SU(3)_c\otimes
SU(3)_L\otimes U(1)_N$ models of Refs.~\cite{pp,mpp}. Using the
notation of Ref.~\cite{rs}, in the entry $({\bf a},{\bf b})_f(N)$,
${\bf a}$ is an irreducible representation of $SU(3)_c$ and ${\bf b}$
is an irreducible representation of $SU(3)_L$. The subindex $f$
means, in an obvious notation, the respective fields of the model and
the second parenthesis contains the value of the $U(1)_N$ generator
when acting on the states in the $({\bf a},{\bf b})$.

Let us consider the 3-3-1 model in which the leptons transform as
$({\bf1},{\bf3^*},-1/3)$ under the 3-3-1 factors:
$(\nu^c_a,\nu_a,l_a)^T_L$ and quarks as in (\ref{modelo1}) (see
Sec.\ref{subsec:341}) . There are
$68$ degrees of freedom.
Left-handed charged leptons and three of the right-handed $d$-type
quarks are in the following representations, three
${\bf6}^*$:
\begin{equation}
{\bf6^*}_{a}=({\bf3}^*,{\bf1})_{d^c_{jL}}(+1/3)+({\bf1},{\bf3}^*)_{a_L}(-1/3),
\label{6l}
\end{equation}
where $a=e,\mu,\tau$, $j=1,2,3$.

Next, two quark generations transforming as $({\bf3},{\bf3},0)$ and
the other two right-handed $d$-type quarks
are in two ${\bf15}$
\begin{equation}
{\bf15}_{Q_{iL}}=({\bf3}^*,{\bf1})_{d^c_{iL}}(+1/3)
+({\bf1},{\bf3}^*)_{X^c_{iL}}(-1/3)+({\bf3},{\bf3})_{Q_{iL}}(0),
\label{151}
\end{equation}
where $i=1,2$; $X^c_{iL}$ are two antitriplets of new left-handed leptons
with charge $(0,0,-1) $. The other quark generation is in one ${\bf20}$
\begin{equation}
{\bf20}_{Q_{3L}}=({\bf1},{\bf1})_{N_{1L}}(0)+({\bf1},{\bf1})_{N_{2L}}(0)
+({\bf3},{\bf3}^*)_{Q_{3L}}(+1/3)+({\bf3}^*,{\bf3})_{Q'_{L}}(-1/3),
\label{152}
\end{equation}
with $N_{iL}$, $i=1,2$ neutral leptons and $Q'_L$ new quarks
trasforming as antitriplets under $SU(3)_c$.
One of the charge $2/3$ quarks are in one ${\bf6}^*$
\begin{equation}
{\bf6}^*_{t'^c_L}=
({\bf3}^*,{\bf1})_{t'^c_{L}}(-2/3)+({\bf1},{\bf3}^*)_{X'^c_{1L}}(+2/3),
\label{r1}
\end{equation}
where $X'^c_{1L}$ is an antitriplet of
extra leptons with $(1,1,0)$ electric charges.
The other 3 charge  2/3 quarks are in three ${\bf6}$
\begin{equation}
{\bf6}_{u_{jR}}=
({\bf3},{\bf1})_{u_{jR}}(+2/3)+({\bf1},{\bf3})_{X'_{jL}}(-2/3),
\label{r3}
\end{equation}
where $X'_{jL}$ are three triplets of leptons with charges
$(1,1,0)$. The quarks $Q'_L$ in (\ref{152}) has their right-handed
components in three ${\bf6}^*$
\begin{equation}
{\bf6}^*_{Q'_{j
R}}=({\bf3},{\bf1})_{Q'_{jR}}(+1/3)+({\bf1},{\bf3})_{X_{3R}}(-1/3).
\label{qp}
\end{equation}

Finally, there are the singlets
of $SU(6)$ corresponding to the usual right-handed charged leptons $l^-_R$,
new leptons $X_{jL}$, $X'_{\alpha L}$ and $N_{iR}$,
where $j=1,2,3; \alpha=1,2,3,4$; $i=1,2$ and $SU(3)_L\otimes U(1)_N$
indices have been omitted.

Maybe before unification, the model must be embedded in a
$(n\!-\!m\!-\!1,\,n>m)$
model. In fact, it is not a trivial issue to show that the
unification in $SU(6)$ may
actually occur~\cite{fn2}. This is so, because in 3-3-1 the couplings
$\alpha_c$
and $\alpha_{3L}$ have $\beta_{c}>\beta_{3L}$. Notice
however that, in order to put all fermion fields of 3-3-1 in $SU(6)$
multiplets, it was necessary to include new fields. It means that
these extra fields have to be added to the minimal 3-3-1 model, and
their effects in the $\beta$-functions must be taken into account.
Therefore, we may have to consider mass threshold corrections for the
$\beta$-functions, since the new particles could have masses below
the unification energy scale, or even, we may not assume the decoupling
theorem. We recall that in the SM with two
Higgs doublets the decoupling theorem~\cite{ac} must not be necessarily
valid, since there are physical effects proportional to
$m^2_{\mbox{Higgs}}$~\cite{dt}. Hence, it could be interesting to study the
way in which the masses of the extra Higgs and exotic
quarks in the model become large, as it has been done in the standard
model scenario for an extended Higgs sector~\cite{dt} or for
the mass difference between fermions of a multiplet~\cite{veltman}.
It means that there is no ``grand
desert'' if 3-3-1 models are realized in nature.

This situation also appears when we consider the embedding of the SM
in 3-3-1. The last model has fields which do not exist in the
minimal SM, but which are in the same multiplet of 3-3-1 with the
known quarks. For instance, the quarks $J$'s have to be added to the
SM transforming as $({\bf3},{\bf1},Q_J)$ under the 3-2-1 factors. The
scalar and vector boson sectors of the SM have also to be extended
with new fields. Hence, we must add
scalar fields transforming as (i) four singlets $({\bf1},{\bf1},Y_S)$:
one with $Y_S=0$, one with $Y_S=1$ and two with $Y_S=2$,
(ii) four doublets $({\bf1},{\bf2},Y_D)$: one with $Y_D=-3$ and three
with $Y_D=1$;  finally, (iii)
one  triplet $({\bf1},{\bf3},-2)$. It is also necessary to add extra
vector bosons ($U^{++},V^+$) which transform as $({\bf1},{\bf2},3)$.
For this reason we believe that 3-3-1 models are not just an
embedding of the SM but an alternative to describe the same
interactions.
\section{Conclusions}
\label{sec:con}
The 3-3-1 models are in fact interesting extensions of the standard
model. They give partial answers to some questions put forward by the
later model and it is also important that new physics could arise at
not too high energies, $\sim 1$ TeV~\cite{pf}. For instance this
sort of models predicts new processes in which the initial states have
the same electric charge as $ff\to W^-V^-$. This processes have only
recently begun to be studied~\cite{rs,pm}.
 Also in some extensions of
these models with spontaneous and/or explicit breaking of $L+B$
symmetry it is possible to have processes with $\vert\Delta L\vert=2$
on kaon decays $k^+\to\pi^-\mu^+\mu^+,\pi^-\mu^+e^+$, and too
$\vert\Delta L\vert=2$ decays of $D$ and $B$ mesons. Experimental
data imply $B(K^+\to \pi^-\mu^+\mu^+)<1.5\times10^{-4}$~\cite{rs}.
The process $e^-e^-\to W^-W^-$ which also could occur in some
extensions of the 3-3-1 models has been recently investigated in other
context~\cite{pm}.

Another interesting
feature of this kind of models is that they include some extensions of
the Higgs sector in the standard model: more doublets, single and
doubly charged, triplets, etc. The scalar sector of these models
deserves detailed study too.

\acknowledgements

I would like to thank the
Con\-se\-lho Na\-cio\-nal de De\-sen\-vol\-vi\-men\-to Cien\-t\'\i
\-fi\-co e Tec\-no\-l\'o\-gi\-co (CNPq) for partial financial support and
M.C. Tijero for reading the manuscript.


\begin{references}

\bibitem{pp} F. Pisano and V. Pleitez, Phys. Rev. D{\bf 46},
410(1992);  P. H. Frampton, Phys. Rev. Lett. {\bf 69}, 2889(1992)
; R. Foot, O.F. Hernandez, F. Pisano and V. Pleitez, Phys.
Rev. D{\bf47}, 4158(1993).
\bibitem{mpp} J.C. Montero, F. Pisano and V. Pleitez, Phys. Rev. D.
{\bf 47}, 2918(1993).
\bibitem{pt} V. Pleitez and M.D. Tonasse, Phys. Rev. D{\bf48},
2353(1993).
\bibitem{pp2} F. Pisano and V. Pleitez, Flavor changing neutral
currents in $SU(3)\otimes U(1)$ models, preprint IFT-P.026/93.
\bibitem{ng} D. Ng, The electroweak theory of $SU(3)\otimes U(1)$,
preprint TRI-PP-92-125.
\bibitem{f} R. Foot, Phys. Rev. D {\bf40}, 3136(1989).
\bibitem{fh} R. Foot and O. F. Hernandez, Phys. Rev. D {\bf41},
2283(1990); Erratum-{\em ibid} D{\bf42}, 948(1990).
\bibitem{s} M. Singer, J.W.F. Valle, and J. Schechter, Phys. Rev.
D{\bf22}, 738(1980); J.W.F. Valle and M. Singer, {\em ibid} D{\bf28},
540(1983).
\bibitem{fn1} We have used the $\bar\lambda$'s for the antitriplets
given in the Appendix of Ref.~\cite{mpp}.
\bibitem{da} D. Amati et al., Nuovo Cimento. {\bf34}, 1732(1964).
\bibitem{voloshin} M.B. Voloshin, Sov. J. Nucl. Phys. {\bf48}, 512(1988).
\bibitem{ppx} That, in fact, this symmetry breaking pattern is possible has
been shown by F. Pisano and V. Pleitez, An $SU(4)_L\otimes U(1)$ electroweak
model, to be submitted for publication.
\bibitem{so} S. Okubo, Phys. Rev. D {\bf16}, 3535(1977).
\bibitem{gs} G. Senjanovic and R.N. Mohapatra, Phys. Rev. D {\bf12},
1502(1975); G. Senjanovic, Nucl. Phys. {\bf B153}, 334(1979).
\bibitem{gelmini} Y. Chikashige, G. Gelmini, R.D. Peccei and M.
Roncadelli, Phys. Lett. {\bf B94}, 499(1980).
\bibitem{gp} H. Georgi and A. Pais, Phys. Rev. D {\bf19}, 2746(1979).
\bibitem{bg} J. Banks and H. Georgi, Phys.\ Rev.\ D {\bf14},
1159(1976).
\bibitem{gg} H. Georgi and S. L. Glashow, Phys. Rev. D
{\bf6}, 429(1972).
\bibitem{rs} R. Slansky, Phys. Rep. C{\bf79}, 1(1981).
\bibitem{fn2} Unified theories based on $SU(15)$ with leptons
($l,l^c,\nu$) and $SU(16)$ with leptons ($\nu,l,l^c,\nu^c$) belonging
to the same multiplet were considered by S.L. Adler, Phys. Lett.
{\bf225B}, 143(1989) and J.C. Pati, A. Salam and J. Strathdee, Nucl.
Phys. {\bf B185}, 445(1981), respectively. Notwithstanding, in these
models the $SU(2)$ subgroup rotating the charged fermion to
their antiparticles includes both lepton and quark sectors.
\bibitem{ac} T.W. Appelquist and J. Carazzone, Phys. Rev. D{\bf11},
2856(1975).
\bibitem{dt} D. Toussaint, Phys. Rev. D {\bf18}, 1626(1978).
\bibitem{veltman} M. Veltman, Nucl. Phys. {\bf B123}, 89(1977).
\bibitem{pf} P.H Frampton, J.T. Liu and B. Charles Rasco, SSC
phenomenology of the 331 model of flavor, hep-ph/9304294.
\bibitem{rs} L.S. Littenberg and R.E. Shrock, Phys. Rev. Lett.
{\bf68}, 443(1992).
\bibitem{pm} C.A. Heusch and P. Minkowski, Lepton flavor violation
induced by heavy Majorana neutrinos, CERN-TH.6606/92.
\end{references}
\end{document}